\documentclass[11pt]{article}
\usepackage{subfigure}
\usepackage{epsfig}

\input epsf
\def\esp{\mathrm{e}}
\def\beq{\begin{eqnarray}}
\def\eeq{\end{eqnarray}}
\def\p{{\bf p}}

\usepackage{amsmath}
\usepackage{amsfonts}
\usepackage{amssymb}

\begin{document}

\begin{titlepage}

\thispagestyle{empty}

\begin{flushright}
{}
\end{flushright}
\vskip 0.9cm

\centerline{\Large \bf  Phase Transitions of Charged Scalars}      
\centerline{\Large \bf  at Finite Temperature and Chemical Potential}

\vskip 0.7cm
\centerline{\large Rachel A. Rosen}

\vspace{0.3cm} 
\centerline{\em Oskar Klein Centre for Cosmoparticle Physics, Department of Physics}
\centerline{\em Stockholm University, AlbaNova SE-10691, Stockholm, Sweden}

\vskip 0.5cm

\begin{abstract}
We calculate the grand canonical partition function at the one-loop level for scalar quantum electrodynamics at finite temperature and chemical potential.  A classical background charge density with a charge opposite that of the scalars ensures the neutrality of the system.  For low density systems we find evidence of a first order phase transition.    We find upper and lower bounds on the transition temperature below which the charged scalars form a condensate.  A first order phase transition may have consequences for helium-core white dwarf stars in which it has been argued that such a condensate of charged helium-4 nuclei could exist.

\end{abstract}

\vspace{0.5in}

\end{titlepage}

\newpage

\section {Introduction and Summary}
There are two ways in which bosons can condense: via Bose-Einstein condensation or via spontaneous symmetry breaking.  Bose-Einstein condensation occurs due to a conserved current associated with the bosons.  Since the total number of particles (or other conserved quantity) can then be fixed, at low temperatures and high densities energy levels become ``overcrowded."   As a result the charge of the system must be stored the zero-momentum ground state.  In theories with spontaneous symmetry breaking, a conserved current is not responsible for the condensation.  Instead, the interactions of the bosons make it energetically favorable for them to condense.

In this paper we consider the first mechanism, Bose-Einstein condensation, in the case that there is both a global $U(1)$ symmetry corresponding to a conserved scalar number density as well as a local gauge symmetry.  In particular, we consider scalar quantum electrodynamics in which the conserved scalar current allows us to introduce a nonzero chemical potential for the scalars.  To ensure the neutrality of the system we consider scalars that reside in a constant background charge density of the opposite charge.  In this work we treat this density as a classical, external background. 

To study the phase transitions of the system, we calculate the grand canonical partition function at the one-loop level.  We find the critical temperature $T_{c1}$ below which the ground state of the system is macroscopically occupied.  Below this temperature some fraction of the scalar charge must reside in a condensate.  We find also an implicit expression for the critical temperature $T_{c2}$ above which all of the scalar charge is stored in thermally excited states.  In between these two temperatures both condensed and uncondensed solutions exist.

For the usual neutral Bose-Einstein condensate these two temperatures coincide.  At the critical temperature $T_c$ the expectation value of the scalars $v$ which is the order parameter of the phase transition goes continuously from zero to nonzero as the temperature of the system is lowered.  This is characteristic of a second order phase transition.  For the charged condensate we find markedly different behavior depending on the relative values of the number density $n$ and the mass $m$.  For low densities, when $n^{1/3} \lesssim \alpha_{\rm em} m$, we find evidence of a first order phase transition: in between the critical temperatures $T_{c1}$ and $T_{c2}$ the value of $v$ jumps discontinuously from zero.  The lower critical temperature $T_{c1}$ coincides with the critical temperature $T_c$ for the neutral condensate, while the upper critical temperature $T_{c2}$ can be several times larger than $T_c$ in the low density regime.  This implies that for the charged condensate, condensation can occur at higher temperatures than for the neutral condensate.  For high densities, when $n^{1/3} \gg \alpha_{\rm em} m$, then $T_{c2}$ approaches $T_{c1}$ and the transition appears to be quite nearly second order.

The latter behavior can be understood through simple arguments.  When the average kinetic energy of a boson is much greater than its Coloumb energy, the gas of bosons can be treated as nearly ideal.  For this to be the case near the transition temperature, it is necessary that $n^{1/3} \gg \alpha_{\rm em} m$.  Thus the physics of the charged condensate in the high density regime should be well approximated by the ideal, neutral condensate.  The main interest of this paper is the low density regime, when $n^{1/3} \lesssim \alpha_{\rm em} m$.  In this regime the mass of the gauge boson in the condensate $m_\gamma \sim e \, v$ becomes comparable with or greater than the transition temperature $T_c$ and a first order phase transition is possible.

Such a system of charged scalars has been considered recently in the context of helium-core white dwarf stars \cite{GGRR1,GGRR3,GGRR4,GGRR5,GGDP}.  In these works helium-4 nuclei play the role of the scalars while a degenerate electron gas is the background charge density.  The number density of the nuclei $n \simeq (0.1~ {\rm MeV})^3$ is such that the average interparticle separation falls between the nuclear and atomic scales.  Thus the nuclei are unable to form neutral atoms, while nuclear effects are negligible.  It is also the case that the number density satisfies $n^{1/3} < \alpha_{\rm em} m$, as $m \simeq 4~{\rm GeV}$.  While standard theory holds that white dwarfs with carbon or oxygen cores will crystalize as they cool, it was argued in Refs. \cite{GGRR3,GGDP} that for helium-core white dwarfs quantum effects will become significant before the crystallization point is reached.  Thus instead of crystallizing, the helium nuclei can condense.

In these works the condensation of helium-4 nuclei was described using a low energy effective field theory.  The work of this paper is relevant to the helium-core white dwarfs if we take scalar QED to be a relativistic effective field theory.  The relativistic theory is not the most appropriate description of the helium-4 condensate.  It is overly restrictive in that it enforces Lorentz invariance - a symmetry we do not expect the low energy system of helium-4 nuclei and electrons to preserve.  In addition it contains a heavy mode which we would expect to be beyond the scope of a low energy theory.  However, it captures the significant features of the condensate including equivalent dispersion relations for the relevant degrees of freedom.

It was recently argued in Ref. \cite{LBGGDP} that, in a neutral system of deuterons and electrons, at sufficiently high densities and low temperatures the spin-1 deuterons will condense.  Such a system may be relevant to certain low-mass brown dwarf stars in which it is expected that such matter can exist in localized regions.  In addition, because deuterons can condense at lower densities than helium, it may be possible to produce such high-density deuteron matter in laboratories in future shock wave compression experiments.  The methods of this paper should be applicable to studying the phase transitions of these systems as well.

\vspace{0.3cm}

Let us make a few comments on the literature.  Gauge theories at finite temperature and symmetry breaking in gauge theories at finite temperature have been studied extensively, early works being \cite{Bernard,DJ,W74} (see also \cite{KapustaGale} and references therein).  In Ref. \cite{KL} spontaneous symmetry breaking in the abelian Higgs model at finite temperature was studied.  It was found that a first order phase transition could take place if $\lambda \ll g^4$ where $\lambda$ is the coupling of the $\phi^4$ interaction and $g$ is the charge of the scalar.  For $\lambda \gtrsim g^4$ the phase transition is second order.  The effect of a nonzero background fermionic current density at zero temperature was also considered in this work.  It was found that an increase of the background charge density could lead to an increase in symmetry breaking.

In Ref. \cite{Linde} the abelian Higgs model with a background fermionic current density at finite temperature was studied in the limit that $\lambda \gg g^4$.  The same model was also considered in Ref. \cite{Kapusta} using a somewhat different method: instead of introducing a background fermionic current, a chemical potential associated with the conserved scalar current was introduced.  The $\lambda \gg g^4$ limit was taken here as well.  Taking this limit allows one to ignore terms that mix the scalar and gauge bosonic fields.  In our work we assume that the $\phi^4$ interaction is negligible compared to the electrostatic interactions.  Thus this mixing term plays a significant role, changing the spectrum of the theory in the condensed phase.  Moreover, the emphasis of these two works is the effect of a conserved current on spontaneous symmetry breaking, rather than the Bose-Einstein condensation of the scalars due to the conserved current.  The latter is the focus of this work.

The effects of nonzero chemical potentials associated with conserved charges on symmetry breaking at finite temperature were also considered in Ref. \cite{HW}.  In Ref. \cite{BBD} an exact expression for the one-loop effective potential for $\phi^4$ theory with a global $U(1)$ charge at finite temperature was obtained.  Refs. \cite{Dolgov,Dolgov2} consider the screening of electric charge and photon polarization in scalar QED at finite temperature and density.  Phase transitions in scalar QED with a nonzero chemical potential at zero temperature due to external magnetic fields were considered in \cite{GGRR5}.

\vspace{0.3cm}

The organization of this work is as follows.  In the following section we briefly review the condensation of charged scalars at zero temperature.  A more thorough discussion can be found in the recent review \cite{GGRR6}.  In section 3 we compute the grand canonical partition function at the one-loop level.  We also treat the UV divergences of the theory.  In section 4 we find the critical temperature $T_{c1}$ below which the scalars have condensed and we find an implicit expression for $T_{c2}$ above which all of the scalars are in thermally excited states.  We plot the expectation value of the scalars $v$ as a function of temperature for various values of the mass $m$ and number density $n$ of the scalars.  We discuss the evidence for a first order phase transition when $n^{1/3} \lesssim \alpha_{\rm em} m $.  In the concluding section we make some comments on the validity of the perturbative expansion and the relevance of our findings to helium-core white dwarf stars.

\section{Charged condensation at zero temperature}
We start by considering the Lagrangian of a charged scalar field $\Phi$ of mass $m$ and a gauge field $A_\mu$ at zero temperature.   The scalars carry charge $g$ which we take to be some multiple of the electric charge $g=Ze$.  We include a background external charge density $g'J_\mu$ which has a charge opposite that of the scalars:
\beq
\label{lagr0}
{\cal{L}} = -\tfrac{1}{4} F_{\mu\nu}^2 +  \vert D_{\mu} \Phi \vert^2
 - m^2 \Phi^{\ast} \Phi -g' A^{\mu} J_{\mu} \, .
\eeq
The covariant derivative for the scalars is defined as $D_\mu \equiv \partial_\mu-i g A_\mu $.

This Lagrangian can arise from a gauge invariant theory in which fermions with charge $g'$ play the role of the background charge density:
\beq
{\cal{L}} = -\tfrac{1}{4} F_{\mu\nu}^2 +  \vert D_{\mu} \Phi \vert^2
 - m^2 \Phi^{\ast} \Phi+ \bar{\psi} (i \gamma^\mu D_\mu-m_F)\psi \, .
\eeq
In this work we consider such a theory in the limit that the fermions are non-dynamical (i.e., frozen in ``by hand" or by some other dynamics).  At distance scales that are greater than the average separation between the fermions, their spatial distribution can be assumed to be uniform. Then, the background charge density can be approximated as $J_\mu \equiv \bar{\psi} \gamma_\mu \psi = J_0 \delta_{\mu 0}$ .

Due to the global $U(1)$ symmetry, there is a conserved scalar current:
\beq
J_\mu ^{\rm{s}} \equiv -i [(D_\mu \Phi)^* \Phi-\Phi^* (D_\mu \Phi)] \, .
\eeq
We can associate a chemical potential $\mu$ with the conserved scalar current.  To incorporate the presence of a conserved current into the Hamiltonian density we make the usual shift ${\cal{H}} \rightarrow {\cal{H}}' = {\cal{H}}-\mu J_0 ^{\rm{s}} $.  For the Lagrangian density the equivalent shift is
\beq
\label{lagr}
{\cal{L}}  \rightarrow {\cal{L}}' = -\tfrac{1}{4} F_{\mu\nu}^2 +  \vert (D_{\mu} - i \mu \delta_{\mu 0}) \Phi \vert^2 - m^2 \Phi^{\ast} \Phi -g' A^{\mu} J_{\mu} \, .
\eeq
For convenience we switch notation and write the scalar field $\Phi$ in terms of a modulus and a phase $\Phi = \tfrac{1}{\sqrt{2}} \sigma\, \esp^{i \alpha}$.  In terms of these variables the Lagrangian density becomes 
\beq
\label{lagr1}
{\cal{L}}'=-\tfrac{1}{4} F_{\mu\nu}^2 + \tfrac{1}{2}(\partial_{\mu}\sigma)^2+
\tfrac{1}{2}(gA_\mu-\partial_\mu \alpha+\mu \delta_{\mu 0})^2 \, \sigma^2- 
\tfrac{1}{2} m^2 \, \sigma^2 - g' A^{\mu} J_{\mu} \, .
\eeq
Written in this form, it is evident that the chemical potential acts as a tachyonic mass for the scalars.  

In these variables the scalar number density is given by 
\beq
\label{J}
 J_0^{\rm{s}} = \,(gA_0-\partial_0 \alpha+\mu) \, \sigma^2 \, .
\eeq
Varying the Lagrangian with respect to $A_\mu$ gives the following equation of motion:
\beq
\label{eom1}
-\partial^\mu F_{\mu \nu} =g(gA_\nu-\partial_\nu \alpha+\mu \delta_{\nu 0}) \sigma^2-g'J_\nu \, .  
\eeq
The system is electrically neutral when the scalar charge density is equal in magnitude to the background charge density:
\beq
g J_0^{\rm{s}} = g' J_0 \, .
\eeq

Varying the Lagrangian with respect to $\sigma$ gives:
\beq
\label{eom2}
\Box \, \sigma = [(gA_\mu-\partial_\mu \alpha+\mu \delta_{\mu 0})^2-m^2] \, \sigma \, ,
\eeq
while varying with respect to $\alpha$ gives the conservation equation for the scalar current:
\beq
\label{J0scalar}
\partial^\mu J_\mu^{\rm{s}} = \partial^\mu \left[(gA_\mu-\partial_\mu \alpha+\mu \delta_{\mu 0}) \, \sigma^2 \right] = 0.
\eeq
Let us take $A_j-\partial_j \alpha = 0$ so that the number density of scalars is constant in time: $J^{\rm{s}}_0 = {\rm const}$.  Subject to this constraint, the equation of motion for the scalars (\ref{eom2}) becomes
\beq
\label{V}
\Box \, \sigma = \frac{(J_0^{{\rm s}})^2}{\sigma^3}-m^2 \sigma\, .
\eeq
This system has a constant, static solution
\beq
\langle \sigma \rangle =\sqrt{\frac{J^{\rm{s}}_0 }{m}} \, .
\eeq
The nonzero expectation value for $\sigma$ indicates that the scalars are in the condensed phase.  It follows from (\ref{J}) that in the condensed phase $\langle g A_0-\partial_0 \alpha \rangle+\mu = m$.  For an electrically neutral system in which the gauge-independent quantity $\langle g A_0-\partial_0 \alpha \rangle$ is equal to zero, condensation occurs when $\mu = m$.  Thus at zero temperature we have a neutral system in which the charged scalars are condensed into a zero-momentum, macroscopic state.  For further discussion of the zero temperature condensate see \cite{GGRR6}.

\section{The Thermodynamic Potential}
In order to study the properties of the charged condensate at finite temperature we start by computing the grand canonical partition function ${\cal Z}$.  We use the functional integral representation of the partition function as it is most suited to the field-theoretic approach adopted above:
\beq
\label{z}
{\cal Z}=N \int [d\Phi] [d\Phi^*] [dA_\mu] \det \left(\frac{\delta F^{\theta}}{\delta \theta} \right) \, \delta(F) \, \exp\left[ \int_0^\beta d\tau \int d^3x {\cal L}'  \right] \, .
\eeq
Here $N$ is an irrelevant normalization constant, $\beta=1/T$ is the inverse temperature and  $\tau = it$ is the imaginary time.  We have introduced a gauge fixing condition $F$ in order to evaluate the functional integral over the gauge fields.  The action is defined by
\beq
S \equiv - \int_0^\beta d\tau \int d^3x {\cal L}' \, ,
\eeq
where the Lagrangian density ${\cal L}'$ is given by (\ref{lagr}) and contains the scalar chemical potential as well as the external current $J_\mu$.  We also make the replacement $\bar{A}_0 = - i A_0$. 

Let us decompose $\Phi$ into real and imaginary parts:
\beq
\Phi = \tfrac{1}{\sqrt{2}} (\phi_1+i\phi_2) \, .
\eeq
The functional integral over the fields is constrained to be periodic so that $\phi(0,{\bf x})=\phi(\beta,{\bf x})$.  Given this constraint, we can Fourier expand the fields as follows:
\beq
\label{p1}
\phi_1(x) =v+ \sqrt{\frac{\beta}{V}} \sum^{\infty}_{n=-\infty} \sum_\p \esp^{i (\omega_n \tau+\p \cdot {\bf x})} \, \phi_{1;n}(\p) \, , \\
\label{p2}
\phi_2(x) = \sqrt{\frac{\beta}{V}} \sum^{\infty}_{n=-\infty} \sum_\p \esp^{i (\omega_n \tau+\p \cdot {\bf x})} \, \phi_{2;n}(\p) \, , \\
\label{A}
A_\mu(x) = \sqrt{\frac{\beta}{V}} \sum^{\infty}_{n=-\infty} \sum_\p \esp^{i (\omega_n \tau+\p \cdot {\bf x})} \, A_{\mu;n}(\p) \, ,
\eeq
where $\omega_n \equiv 2 \pi n T$ and $V$ is the volume of the system.  For the field $\phi_1$ we have separated out a constant part $v$ which is independent of ${\bf x}$ and $\tau$ so that $\phi_{1;n=0}(\p=0) = 0$.  Thus $v$ represents the thermal average of the field: $\langle \phi_1(x) \rangle = v$.  We do this in anticipation of the condensation of the scalars into the $n=0$, $\p=0$ state.  The partition function will be at a minimum with respect to the free parameter $v$:
\beq
\frac{\partial \ln {\cal Z}}{\partial v} =0 \, .
\eeq
Without loss of generality we can set  $\langle \phi_2(x) \rangle = 0$.  

One could also separate out an expectation value for the scalar potential:  $\langle A_0(x) \rangle = a_0$.  The partition function will also be at a minimum with respect to $a_0$:
\beq
\label{Za}
\frac{\partial \ln {\cal Z}}{\partial a_0} =0 \,.
\eeq
It can be seen from the Lagrangian (\ref{lagr1}) that a nonzero expectation value for the scalar potential acts as a shift in the chemical potential, except in its coupling to the external current $J_0$.  Thus equation (\ref{Za}) is equivalent to 
\beq
\label{Zmua}
g\frac{1}{\beta V}  \frac{\partial \ln {\cal Z}}{\partial \mu} -g' J_0=0 \, .
\eeq
The first term on the right hand side is the scalar charge density:
\beq
\label{Zmu}
\frac{1}{\beta V} \frac{\partial \ln {\cal Z}}{\partial \mu}= \langle J_0^{\rm s} \rangle \, .
\eeq
Thus minimization of $\ln {\cal Z}$ with respect to $a_0$ gives the condition of charge neutrality: $g \langle J_0^{\rm s} \rangle = g' J_0$.

Equation (\ref{Zmua}) depends only on the total effective chemical potential, i.e., on the sum  $\mu_{\rm eff} = \mu+ga_0$ and not on $a_0$ alone.  Thus minimization does not fix the value of $ a_0$ independently of $\mu$.  Instead, equation (\ref{Zmu}) can be used to determine $\mu_{\rm eff}$, given a fixed charge density.  One could consider a configuration with a nonzero $a_0$ acting as an external chemical potential for the scalars, corresponding to some uncompensated charge on a surface at infinity.  Such a setup was studied in earlier works \cite{GGRR1,GGRR2} in the zero-temperature limit.  For simplicity, we will assume a configuration with no surface charge and take $a_0 = 0$ in what follows.\footnote{To consider a system with a nonzero surface charge, one need only make the replacement $\mu \rightarrow  \mu_{\rm eff}$ in the following expressions.  The results are otherwise unaffected.}

Terms in the Lagrangian that are linear in the excitations of the fields above their background values will contribute to the action an amount proportional to $\phi_{1,2;n=0}(\p=0)$ or $\bar{A}_{0;n=0}(\p=0)$ for the scalar and gauge fields respectively.  Thus given the above definitions, these linear terms in the Lagrangian will not contribute to the partition function.  Terms that are linear in $A_{j;n}(\p)$ do not appear in the Lagrangian as long as the external current $J_j$ is zero.

\vspace{0.3cm}

In order to compute the functional integral we must first choose a gauge $F$.  The most natural choice is the unitary gauge in which the phase of the scalar field is set to zero: $F = \alpha = 0$.  In this gauge the physical content of the theory is apparent and no unphysical degrees of freedom need to be introduced.  However, the unitary gauge is known to give incorrect results at the one-loop level.  This is attributed to the fact that Lagrangian in the unitary gauge does not correspond to a renormalizable theory \cite{DJ,W74}.  We choose instead a renormalizable $R_\xi$ gauge.
A family of covariant gauges is given by
\beq
F = \partial_\mu A^\mu+gv\phi_2 -f(x)=0\, ,
\eeq
for some arbitrary function $f(x)$.  With this choice the partition function (\ref{z}) becomes
\beq
\label{z2}
{\cal Z} = N \int [d\phi_1] [d\phi_2] [dA_\mu] \det \left(\Box+ g^2v^2\right) \, \delta(\partial_\mu A^\mu+gv\phi_2-f(x)) \, \esp^{-S} \, .
\eeq
The delta function in the partition function can be incorporated into the Lagrangian via
\beq
\label{lagr2}
{\cal L}' \rightarrow {\cal L}' -\frac{1}{2 \xi}(\partial_\mu A^\mu+gv\phi_2)^2 \, .
\eeq
In what follows we will keep the gauge parameter $\xi$ general in order to check that our results are independent of the gauge fixing condition.

\vspace{0.3cm}

To compute the partition function we proceed via the mean field approximation.  We take the field expansions (\ref{p1}), (\ref{p2}) and (\ref{A}) and substitute them into the Lagrangian (\ref{lagr2}).  We expand the Lagrangian to second order in the fields, which we assume to be small fluctuations above the mean field values.  We neglect the terms that are linear in the fields, as per the discussion above.  For convenience we split the gauge boson into transverse and longitudinal components, the transverse components being given by
\beq
A_j^{tr} = A_j-\frac{\partial_j}{\nabla^2} (\partial_k A_k) \, .
\eeq
Upon integrating, the action (to second order) can be written as a sum of the tree-level component, the contribution coming from the transverse photons and a contribution coming from the scalars and remaining photon degrees of freedom:
\beq
S_2 = S_0+S_{tr}+S_s \, .
\eeq
The tree-level action is given by
\beq
S_0=\tfrac{1}{2} \beta V (m^2-\mu^2) v^2 \, .
\eeq
The contribution to the action coming from the transverse components of the gauge boson is
\beq
S_{tr} = \tfrac{1}{2}\beta^2 \sum_{n} \sum_\p \sum_{i=1,2} A_{i;-n}^{tr}(-\p) (\omega_n^2+\p^2 + g^2 v^2) A_{i;n}^{tr}(\p) \, , \\
\eeq
where the sum over $i$ is the sum over both transverse degrees of freedom.  These components acquire a mass when the scalar is condensed, i.e., when $v \neq 0$.

The remaining scalar and photon degrees of freedom mix with each other.  Their contribution to the action is given by
\beq
S_s=\tfrac{1}{2} \beta^2 \sum_{n} \sum_\p \left(\phi_{1;-n}(-\p),\phi_{2;-n}(-\p),\bar{A}_{0;-n}(-\p),A_{l;-n}(-\p) \right) D 
\left( \begin{array}{c}\phi_{1;n}(\p) \\ \phi_{2;n}(\p) \\ \bar{A}_{0;n}(\p) \\ A_{l; n}(\p) \end{array} \nonumber \right) ,\\
\eeq
where  $D= $
\beq
\label{D}
\left(
\begin{array}{cccc}
\omega_n^2+\p^2+m^2-\mu^2  & -2 \mu \omega_n & -2igv \mu  & 0 \\
+2 \mu \omega_n  & \omega_n^2+\p^ 2 +m^2-\mu^2+\tfrac{1}{\xi} g^2 v^2 & +igv\omega_n (1- \tfrac{1}{\xi}) & +igv|\p|  (1- \tfrac{1}{\xi})\\
-2igv \mu &  -igv\omega_n (1- \tfrac{1}{\xi}) & \tfrac{1}{\xi} \omega_n^2+\p^2 + g^2 v^2 & -\omega_n |\p| (1- \tfrac{1}{\xi}) \\
0 & -igv|\p|  (1- \tfrac{1}{\xi}) &  -\omega_n |\p| (1- \tfrac{1}{\xi})  &  \omega_n^2+ \tfrac{1}{\xi}\p^2 + g^2 v^2 \nonumber
\end{array} \right) . \\ 
\eeq
Let us make a few comments on this matrix.  In the decoupling limit $g \rightarrow 0$ the scalar degrees of freedom decouple from the photon degrees of freedom as expected.  In this limit we recover the theory for the neutral condensate together with decoupled, massless photons.  When $g \neq 0$ but the scalar field is not condensed, i.e., when $v=0$, the same decoupling takes place.  In the uncondensed phase the partition function for the charged scalars is the same as that for the neutral scalars and decoupled photons, at the one-loop level. 

An important element of this matrix is the term that mixes the scalar field $\phi_1$ with the $0$-component of the gauge field: $2igv \mu \phi_1 \bar{A}_0$.  In the absence of a chemical potential for the scalars (i.e., when $\mu = 0$), an appropriate choice of gauge will decouple the gauge degrees of freedom from the scalars, even when $v$ is nonzero.  When $\mu \neq 0$ this decoupling no longer occurs, because of this term.  Due to this mixing, the spectrum of the theory in the condensed phase is significantly different from that of the neutral condensate.

\vspace{0.3cm}

Carrying out the functional integrations, the logarithm of the partition function (\ref{z2}) can be expressed as a sum of the various components
\beq
\ln {\cal Z} = \ln  {\cal Z} _0+ \ln {\cal Z}_{tr}+ \ln  {\cal Z}_{s}+ \ln  {\cal Z}_{gh} \, .
\eeq
There is also a constant term coming from the overall normalization which is temperature independent and which we can neglect.

The first term comes from the tree-level action
\beq
\ln  {\cal Z}_0 = -\tfrac{1}{2} \beta V (m^2-\mu^2) v^2  \, , 
\eeq
the second term from the two transverse photon polarizations
\beq
\ln  {\cal Z}_{tr}= -\tfrac{1}{2}(2) \beta^2 \sum_{n} \sum_\p \,  \ln \det (\omega_n^2+\p^2 + g^2 v^2) \, , 
\eeq
and the third term from the scalar and remaining photon degrees of freedom
\beq
\label{zs}
\ln  {\cal Z}_{s}=-\tfrac{1}{2} \beta^2\sum_{n} \sum_\p \, \ln (\xi \det D)  \, .
\eeq
The last term arises from the ghost determinant in the partition function (\ref{z2}). 
\beq
\ln  {\cal Z}_{gh}=\beta^2 \sum_{n}  \sum_\p \, \ln \det (\omega_n^2+\p^2 + g^2 v^2) \, .
\eeq
Formally, the ghost term exactly cancels the contribution coming from the transverse degrees of freedom.  This allows us to simplify our calculations.

The determinant of $D$ can be written in the following  form
\beq
\label{det}
\xi \det D =  (\omega_n^2+ \omega_+^2)(\omega_n^2+ \omega_-^2)(\omega_n^2+ \omega_1^2)(\omega_n^2+ \omega_2^2) \, ,
\eeq
where the dispersion relations $\omega_+$, $\omega_-$, $\omega_1$ and $\omega_2$ are functions of $\p,m,\mu,gv,$ and $\xi$.  We will denote them by $\omega_\alpha$.  Due to their length we will not write the full expressions here.  Below we will give their exact expressions when evaluated on the solutions to the tree-level equations of motion for the scalar.  We will also find expressions for when the tree-level scalar equations of motion are not satisfied.

This form for $\xi \det D$ allows us to carry out the sum over $n$ in (\ref{zs}) (see \cite{Bernard} for more details).  We replace the sum over $\p$ by the appropriate integral.  The thermodynamic potential $\Omega = -\tfrac{1}{\beta V} \ln  {\cal Z}$ is then:
\beq
\label{Om}
\Omega= \tfrac{1}{2} (m^2-\mu^2) v^2+\sum_\alpha \int{\frac{d^3p}{(2 \pi)^3}}  \tfrac{1}{2}  \omega_\alpha
+\sum_\alpha \frac{1}{\beta} \int{\frac{d^3p}{(2 \pi)^3}} \ln \left(1-\esp^{- \beta \omega_\alpha} \right) \, .
\eeq
The first term is the tree-level potential.  The second term is due to the zero point energies of the fields.  The third term comes from the thermal excitations of the fields.

\vspace{0.3cm}

The thermodynamic potential is subject to two constraints.  First, it should be at a minimum with respect to the free parameter $v$:  
\beq
\label{con1}
\frac{\partial \Omega}{\partial v}=0 \, .
\eeq
For the tree-level potential $\Omega_{\rm tree}= \tfrac{1}{2} (m^2-\mu^2) v^2$ we see that the first condition is satisfied in two ways: in the condensed phase when $v \neq 0$ and $\mu = m$ or in the uncondensed phase when $v=0$ and $\mu \neq m$.  We will refer to these solutions as the ``on-shell" solutions since they satisfy the tree-level equations of motion for the scalar.

Second, the thermodynamic potential should give the fixed number density of scalars when differentiated with respect to $\mu$:
\beq
\label{con2}
n \equiv \langle J_0^{\rm s} \rangle = -\frac{\partial \Omega}{\partial \mu}\, .
\eeq
When applied to the tree-level potential, this constraint gives
\beq
n= -\frac{\partial \Omega_{\rm tree}}{\partial \mu} = \mu v^2 \, .
\eeq
In the condensed phase this is satisfied when $v=\sqrt{n/m}$, as we found in the previous section for the condensate at zero temperature.  In the uncondensed phase ($v=0$) this expression cannot be satisfied.  This is unsurprising as we have neglected contributions to the number density coming from thermal fluctuations.  In the absence of these fluctuations, i.e. at zero temperature, the charge of the system must be stored in the condensate.

\vspace{0.3cm}

The dispersion relations $\omega_\alpha$ simplify greatly when evaluated on the on-shell solutions.  For the uncondensed phase we set $v=0$ in the determinant of $D$ (equation (\ref{D})).  Then $\det D$ can be factored according to equation (\ref{det}).  We find:
\beq
\label{ov0}
\omega_\pm = \sqrt{\p^2+m^2} \pm \mu \, , ~~~~ \omega_{1,2}  =  |\p| \, .
\eeq
The first two dispersion relations are for the scalar particle and antiparticle and are the same as those for a neutral system at finite chemical potential.  The second two are for the photon degrees of freedom.  In the uncondensed phase, the spectrum of the theory and thus the one-loop thermodynamic potential is the same as for neutral scalars and decoupled, massless photons as mentioned above.

To find the dispersion relations in the condensed phase we set $\mu = m$ in the determinant of $D$.  This gives
\beq
\label{opm}
\omega_\pm^2 = \p^2+\tfrac{1}{2} g^2 v^2+2m^2 \pm \sqrt{4 m^2 \p^2+(2 m^2-\tfrac{1}{2}g^2 v^2)^2}\, ,
\eeq
\beq
\label{o12}
\omega_{1,2}^2  = \p^2+g^2 v^2 \, .
\eeq
These dispersion relations agree with those found for the charged condensate at zero temperature in \cite{GGRR1}, with the photon mass $m_\gamma = g \sqrt{n/m}$ replaced by the more general $g v$.  In the condensed phase the first two dispersion relations (\ref{opm}) no longer have the simple interpretation of corresponding to the scalar particle and antiparticle.  In the condensate the gauge symmetry is broken.  The gauge boson becomes massive by eating one of the scalar degrees of freedom.  Thus we can think of these two dispersion relations as corresponding to the longitudinal component of the photon and the remaining scalar degree of freedom, though in reality they correspond to a linear combination of the scalar and gauge fields.  Their masses are found by taking $\p=0$: $\omega_\pm(\p=0)=2m,~gv$.  Both modes are massive.  Thus unlike the neutral condensate which contains a massless particle in the condensed phase, the charged condensate has a mass gap.

For $m \gg gv $ the relations (\ref{opm}) simplify to
\beq
\omega_+ \simeq \sqrt{\p^2+m^2}+m \, ,
\eeq
and
\beq
\omega_-^2 \simeq g^2 v^2+\frac{\p^2(\p^2-g^2v^2)}{4m^2} \, , ~~ \p^2 \ll 2 g v m \, ,\\
\omega_- \simeq \sqrt{\p^2+m^2}-m \, , ~~ \p^2 \gg 2 g v m \, .
\eeq

\vspace{0.3cm}

The solutions to equation (\ref{con1}) calculated from the tree-level potential will be modified by finite temperature effects.  In particular, we will find that at non-zero temperature, $\mu = m$ no longer holds identically in the condensed phase.  Moreover, in order to solve the second constraint equation (\ref{con2}) at finite temperature, we need to know the dispersion relations $\omega_\alpha$ as a function of $\mu$, away from $\mu =m$.  Thus to understand the full thermodynamic potential and its constraints, we must generalize the $\omega_\alpha$ given above to the case that the tree-level solutions to the scalar equations of motion are not satisfied, when $v \neq 0$ and $\mu \neq m$.

To compute the thermodynamic potential we have used the standard background field method.  For gauge theories this method gives a unique result only when the background field is a solution to the tree-level equations of motion.  If this condition is not satisfied, the background field method can give gauge dependent results for physical quantities.  This effect can be seen here by considering the matrix $D$ given in (\ref{D}).  Calculating the determinant of $D$, one finds terms that depend on the gauge fixing parameter in the form $\xi (m^2-\mu^2) g^2 v^2(\ldots)$.  These terms vanish on-shell, i.e., when either $v = 0$ or $\mu = m$.  Thus the dispersion relations found above are independent of $\xi$ as we would expect.  However, when $v \neq 0$ and $\mu \neq m$ the determinant of $D$ and thus the more general dispersion relations become dependent on the gauge fixing condition.

In order to determine the general, gauge-independent dispersion relations from $\det D$ we would first have to find the unique off-shell potential.  For our purposes, however, it is possible to determine the more general $\omega_\alpha$, given our knowledge of their behavior when $v = 0$ or $\mu = m$, in combination with constraints coming from a Ward identity.  We will do this now.

The photon degrees of freedom do not contribute to the number density of the scalars.  Thus the generalized dispersion relations $\omega_{1,2}$ must be independent of $\mu$.  It follows that they are the same when $\mu \neq m$ as when $\mu = m$:
\beq
\omega_{1,2}^2  = \p^2+g^2 v^2 \, .
\eeq
To find the dispersion relations for the scalar degrees of freedom, we start by parametrizing $\omega_\pm^2$ by 
\beq
\label{gen}
\omega_\pm^2 = A \pm \sqrt{B} \, .
\eeq 
By requiring that the general dispersion relations reduce to those found above when $v=0$ or when $\mu = m$ we find 
\beq
A &=& \p^2+m^2+\mu^2+\frac{1}{2} g^2 v^2 +a(m,gv,\mu) \frac{m^2-\mu^2}{m^2}g^2 v^2 \, , \nonumber \\ 
B&=& 4\p^2 \mu^2 + 4m^2 \mu^2-2m^2g^2v^2+\frac{1}{4}g^4 v^4 +b(m,gv, \mu)(m^2-\mu^2)g^2 v^2\, .
\eeq
The functions $a$ and $b$ are not fixed by this requirement.  They can in general represent a series expansion in $m^2 - \mu^2$ and $g^2v^2$ with undetermined coefficients.  We take $a$ and $b$ to be independent of $|\p|$, in order to maintain the appropriate high-momentum behavior of the theory.

For the neutral condensate, $\mu = m$ identically in the condensed phase, even at finite temperature.  For the charged condensate this is not necessarily the case.  This is due to the fact that for the charged condensate the dispersion relations depend on $v$.  However, any deviation of $\mu$ away from $m$ should be suppressed by $\alpha_{\rm em}$ and vanish in the limit that $g \rightarrow 0$.  Thus for the low density systems that are the primary interest of this paper, the $a$ and $b$ terms are subdominant as they are $\alpha_{\rm em}$-suppressed compared to the leading order terms in $A$ and $B$.  For high density systems they can be relevant, when $gv \gg m$.

A Ward identity can be used to fix $a$ and $b$.  The sum over the zero point energies in the thermodynamic potential (\ref{Om}) is UV divergent and thus must be renormalized.  Because of the Ward identity associated with the conserved scalar current, one can show that the conserved current is not subject to infinite renormalization, even in cases of spontaneous symmetry breaking (see, e.g., [19]).  It follows that the divergences in the potential should be independent of $\mu$.  Expanding the dispersion relations (\ref{gen}) for large $p$ we find:
\beq
 \int{\frac{d^3p}{(2 \pi)^3}} ( \omega_+ +\omega_-) \simeq |\p| \, \left[2+ \frac{f_2(m,gv,\mu)}{\p^2}+\frac{f_4(m,gv,\mu)}{\p^4}+O\left(\frac{1}{\p^6}\right)\right]  \, ,
\eeq
where $f_2$ and $f_4$ depend on $a$ and $b$.  The three terms on the r.h.s. represent the quartic, quadratic and logarithmic divergences of the scalar sector.  These divergences can be made independent of $\mu$ for appropriate choices of $a$ and $b$.  This uniquely fixes $a=0$ and $b=-2$.  The generalized dispersion relations are thus:
\beq
\label{oalpha1}
\omega_\pm^2 &=& \p^2+m^2+\mu^2+\frac{1}{2} g^2 v^2 \pm
\sqrt{4\p^2 \mu^2 + 4m^2 \mu^2+2g^2 v^2 \mu^2-4g^2v^2m^2+\frac{1}{4}g^4 v^4} \, . \nonumber \\
\eeq

Let us now treat the remaining, $\mu$-independent divergences in the thermodynamic potential.  To regularize the divergent terms we introduce an ultraviolet cutoff $\Lambda_c$ and take $\vert \p \vert < \Lambda_c$.  After integration, and in the limit of large $\Lambda_c$, the divergent contributions to the thermodynamic potential coming from both the scalar and gauge degrees of freedom are
\beq
\sum_\alpha \int \frac{d^3p}{(2 \pi)^3}  \, \tfrac{1}{2}  \omega_\alpha&=&
  \left[\frac{\Lambda_c^4}{4 \pi^2}+\frac{m^2 \Lambda_c^2}{8 \pi^2}-\frac{m^4}{16 \pi^2}\ln\left(\frac{2\Lambda_c}{m}\right) \right] \nonumber \\
 &&+\left[\frac{3 \Lambda_c^2}{16 \pi^2}+\frac{3 m^2}{16 \pi^2} \ln \left(\frac{2\Lambda_c}{m}\right) \right] g^2 v^2 \nonumber \\
 &&-\left[\frac{1}{32 \pi^2}\ln\left(\frac{2\Lambda_c}{m}\right)+\frac{1}{16 \pi^2}\ln\left(\frac{2\Lambda_c}{gv}\right)\right] g^4 v^4 +{\rm finite~ terms}\, . \nonumber \\
\eeq
The first term renormalizes the vacuum energy density.  This term is independent of temperature and can be trivially subtracted.  The second term renormalizes the mass of the scalar and the third term renormalizes the coupling of the $\phi^4$ interaction, had we included one at the start.  We can add to the Lagrangian counterterms of the form
\beq
{\cal L}_{\rm c.t.} = \delta \Lambda-\delta m^2 \Phi^{\ast} \Phi -\delta \lambda (\Phi^{\ast} \Phi)^2 \, ,
\eeq
to absorb these divergences.  

To renormalize the zero-temperature thermodynamic potential to leading order in $\alpha_{\rm em}$, we can evaluate the contributions from the zero-point energy on the tree-level solution $\mu = m$.  We also, for the moment, ignore the divergences that multiply $v^4$ as they are also higher order in $\alpha_{\rm em}$.  We choose our renormalization conditions so that at zero temperature the thermodynamic potential is finite and is at a minimum when with respect to $v$ when $\mu=m$. This fixes $\delta \Lambda$ and $\delta m^2$ so that
\beq
\Omega_{\rm tree}+\Omega_{\rm z.p.}+\Omega_{\rm c.t.} =  \tfrac{1}{2} (m^2-\mu^2) v^2 +O(\alpha_{\rm em}^2) \, .
\eeq

The presence of a $v^4$ term in the potential will shift the zero-temperature values of $v$ and $\mu$.  Moreover, the dependence of the divergent term that multiplies $v^4$ on the logarithm of $v$ means that the coupling of the $v^4$ interaction will run mildly with temperature.  However this term is suppressed by $\alpha_{\rm em}$ compared to both the other zero temperature terms and the finite temperature terms.  We will neglect its contribution to the potential when studying finite temperature effects in what follows.

\section{Phase Transitions}
In section 3 we found that the minimization of the tree-level potential with respect to $v$ has two solutions: one when $v=0$ and one for arbitrary $v$ when $\mu=m$.  We also observed that at zero temperature the uncondensed solution $v=0$ cannot satisfy the constraint equation (\ref{con2}).  At zero temperature all of the scalars reside in the condensate.  More generally, there exists a critical temperature $T_{c1}$ below which some fraction of the conserved scalar charge must be in the condensed phase.  Likewise, there exists a critical temperature $T_{c2}$ above which $v=0$ is the only solution to both equations (\ref{con1}) and (\ref{con2}).  Between these two temperatures the scalar field will undergo a phase transition into the condensed state.  In this section we will determine $T_{c1}$ and $T_{c2}$ by applying the constraints (\ref{con1}) and (\ref{con2}) to the full thermodynamic potential.  We will also discuss features of the phase transition.

Let us consider the nonrelativistic limit $n^{1/3} \ll m$.  In this case $\omega_+ \simeq 2m$.  Since $gv \ll m$ in this limit, the $\omega_+$ term in the thermodynamic potential is exponentially suppressed compared the $\omega_-$ and $\omega_{1,2}$ terms.  Thus we can neglect its contribution:
\beq
\label{Om2}
\Omega \simeq \tfrac{1}{2} (m^2-\mu^2) v^2+T\int{\frac{d^3p}{(2 \pi)^3}} \ln \left(1-\esp^{- \beta \omega_-} \right) +2\,T\int{\frac{d^3p}{(2 \pi)^3}} \ln \left(1-\esp^{- \beta\sqrt{\p^2+g^2v^2}} \right) .
\eeq
The thermodynamic potential should be at a minimum with respect to $v$: $\partial \Omega/\partial v=0$.  Since $v$ appears in the dispersion relations only in the form $g^2 v^2$ we can write this condition as
\beq
\label{Omdv}
(m^2-\mu^2) v+2g^2 v \int{\frac{d^3p}{(2 \pi)^3}} \left(\frac{\partial \omega_-}{\partial (g^2 v^2)} \frac{1}{\esp^{\beta \omega_-}-1} +\frac{1}{\sqrt{\p^2+g^2v^2}} \frac{1}{\esp^{\beta\sqrt{\p^2+g^2v^2}}-1}\right) =0\, . \nonumber \\
\eeq
This equation has two solutions.   The first solution is the same as for the tree-level potential, when $v=0$. The other solution is given by 
\beq
\label{sol2}
\mu^2=m^2+2g^2 \int{\frac{d^3p}{(2 \pi)^3}} \left(\frac{\partial \omega_-}{\partial (g^2 v^2)} \frac{1}{\esp^{\beta \omega_-}-1}+\frac{1}{\sqrt{\p^2+g^2v^2}} \frac{1}{\esp^{\beta\sqrt{\p^2+g^2v^2}}-1}\right)\,.
\eeq
At finite temperature, $\mu$ is shifted away from its tree-level value $m$ by an amount proportional to $\alpha_{\rm em}$.  It is straightforward to see that $\omega_-$ as given by (\ref{oalpha1}) increases monotonically as a function of $g^2v^2$, i.e., $\partial \omega_-/\partial(g^2v^2) \geq 0$ for arbitrary $v$.  Thus at finite temperature, the second solution to $\partial \Omega/\partial v=0$ requires $\mu > m$.  

For the neutral condensate, the integral in the thermodynamic potential corresponding to the $\omega_-$ mode is convergent only when $\mu \leq m$.  For the charged condensate however, when $v \neq 0$, solutions with $\mu > m$ are possible because of the contribution $gv$ makes to the mass of this mode.  Still, when $v=0$ we must have $\mu \leq m$, as in the neutral case.   

It follows that $v=0$ is not a solution to the above equation (\ref{sol2}).  The absence of a second solution at $v=0$ implies that $v$ must change discontinuously when going from the uncondensed phase to the condensed phase.  This is indicative of a first order phase transition.  We will see that this discontinuity is more pronounced in the low density regime.

\vspace{0.3cm}

Let us consider the $v=0$ solution.  The general dispersion relations given by (\ref{oalpha1}) reduce to those found above in (\ref{ov0}).  In particular
\beq
\omega_- = \sqrt{\p^2+m^2}-\mu \, .
\eeq

Differentiating the potential with respect to $\mu$ gives the usual result for number density:
\beq
n=-\frac{\partial \Omega}{\partial \mu} =\int{\frac{d^3p}{(2 \pi)^3}}\frac{1}{\esp^{\beta (\sqrt{\p^2+m^2}-\mu)}-1}\, .
\eeq
All of the particles are in thermally excited states.  Fixing $n$ gives an implicit expression for the chemical potential as a function of temperature.  The critical temperature $T_{c1}$ is defined to be the minimum temperature at which all the particles are still in excited states.  It is found by taking $\mu = m$ and solving for $T$.  The result is the same as for the critical temperature of the neutral condensate:
\beq
\label{Tc1}
T_{c1}  = \frac{2 \pi}{m}\, \left( \frac{n}{\zeta(3/2)}\right)^{2/3} ~~~~ {\rm{for }} ~ n^{1/3} \ll m\, .
\eeq
The temperature $T_{c1}$ puts a lower bound on the condensation temperature.  Below this temperature no $v=0$ solution exists.  However, unlike in the neutral case, it may be possible for the charged scalars to condense before $T_{c1}$ is reached, as we will now discuss.

\vspace{0.3cm}

When $v \neq 0$ the number density is given by
\beq
\label{n}
n = \mu v^2-\int{\frac{d^3p}{(2 \pi)^3}} \frac{\partial \omega_-}{\partial \mu} \frac{1}{\esp^{\beta \omega_-}-1} \, ,
\eeq 
where $\omega_-$ is given by (\ref{oalpha1}).  The first term on the right hand side represents the fraction of particles in the condensate while the second term gives the particles that remain in thermally excited states.  This equation combined with (\ref{sol2}) gives an implicit expression for the value of the condensate $v$ as a function of temperature for fixed $n$.  As noted above, the finite temperature correction to $\mu^2 -m^2$ is suppressed by $\alpha_{\rm em}$.  As a lowest order approximation we can evaluate the above expression (\ref{n}) on the tree-level solution $\mu = m$ to find $v(T)$.

\begin{figure}[t]
\begin{center}
\subfigure[Nonrelativistic regime: $n^{1/3} < m$]{\epsfig{figure=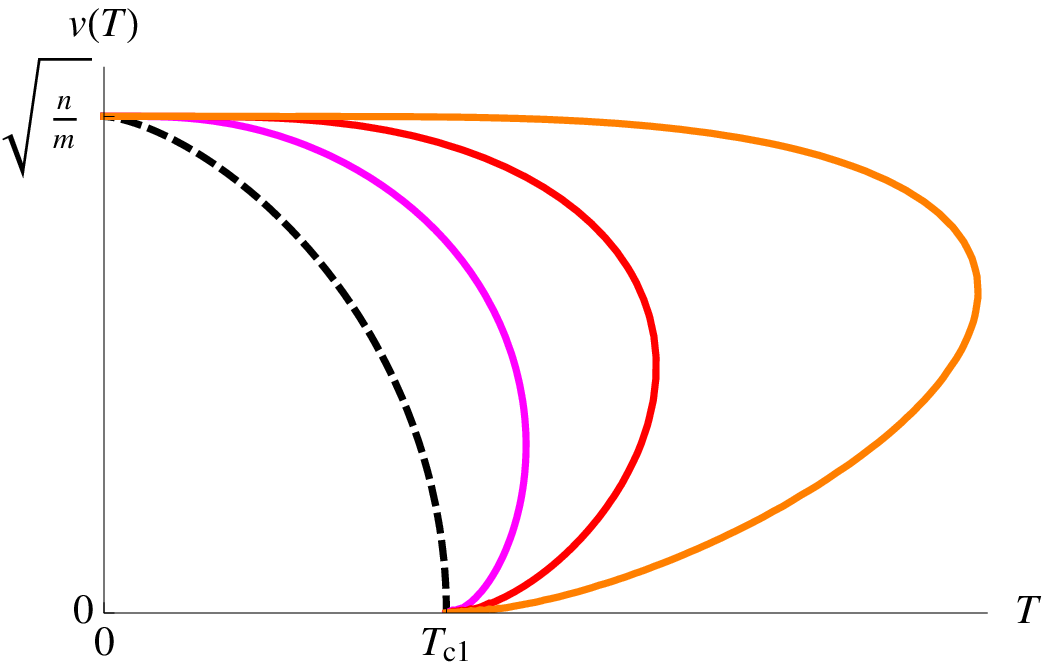,width=.45\textwidth}}
\hskip .15in
\subfigure[Relativistic regime: $n^{1/3} \gtrsim m$]{\epsfig{figure=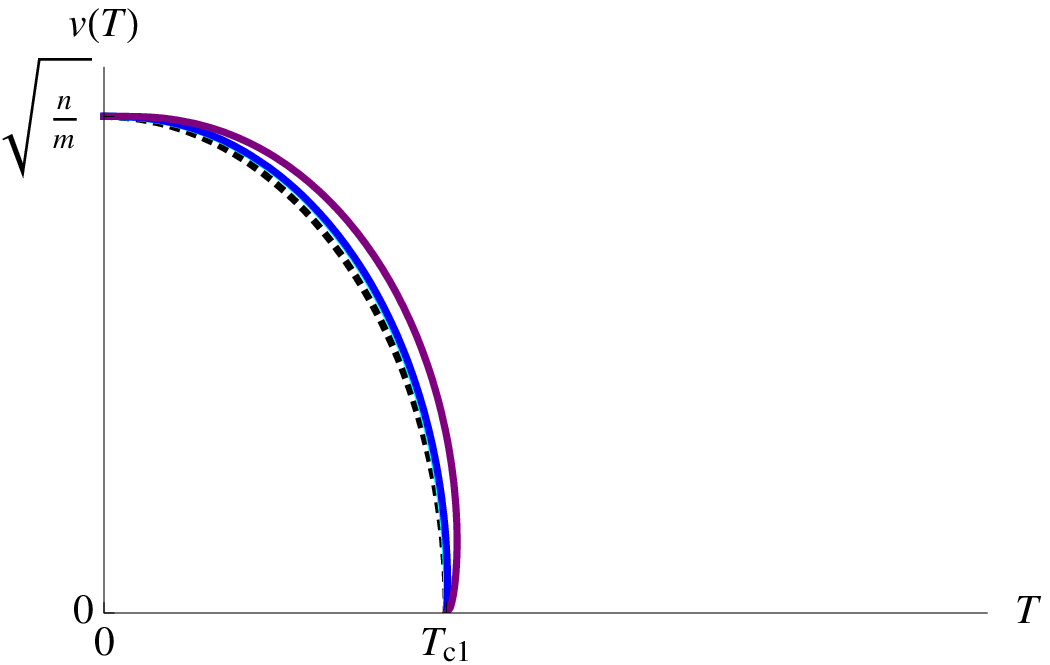,width=.45\textwidth}}\\
\end{center}
\caption{The scalar expectation value $v$ as a function of temperature for various values of $n^{1/3}/\alpha_{\rm em}m$ when $\mu =m$ and $g=2e$.}
\label{fig1}
\end{figure}

In Figure 1 we plot $v$ as a function of temperature for various values of $n^{1/3}/\alpha_{\rm em} m$ using expression (\ref{n}) and taking $\mu = m$.  In the relativistic regime we also include the contribution to number density coming from the $\omega_+$ mode.  We take the charge of the scalars to be $g= 2e$ with a system of helium nuclei in mind.  The values of $v$ at $T=0$ and of the critical temperature $T_{c1}$ are of course different for different values of $m$ and $n$, but we rescale the results for easy comparison.  

In Figure 1a we plot $v(T)$ in the nonrelativistic regime when $n^{1/3} < m$.  The orange line corresponds to $n^{1/3}=1/10 \,  \alpha_{\rm em} m$, the red line to $n^{1/3}= \alpha_{\rm em} m$, and the pink line to $n^{1/3}=10\,  \alpha_{\rm em} m$.  For comparison we have also plotted $v(T)$ for the neutral condensate in the nonrelativistic limit.  This is the black dashed line.  In Figure 1b we plot $v(T)$ in the relativistic regime when $n^{1/3} \gtrsim m$.  The purple line corresponds to $n^{1/3}=10^2\, \alpha_{\rm em}m$, the dark blue line to $n^{1/3}=10^3 \,\alpha_{\rm em}m$ and the light blue line to $n^{1/3}=10^4 \, \alpha_{\rm em} m$.  The black dotted line is $v(T)$ for the neutral condensate in the relativistic limit.  All of the curves share the solution $v=0$ for $T>T_{c1}$.

The shape of the contours for $n^{1/3} \lesssim  \alpha_{\rm em} m$ is characteristic of a first order phase transition.  At sufficiently high temperatures the only solution is $v=0$.  But as the temperature drops, at some temperature $T_{c2} > T_{c1}$, the $v(T)$ curve ceases to be single-valued.  Both the uncondensed solution and the condensed solution exist simultaneously.  Moreover, in between $T_{c2}$ and $T_{c1}$ the value of $v$ must change discontinuously from zero as the system cools.   At temperatures below $T_{c1}$ only the condensate solution exists.  

The temperature $T_{c2}$ is the maximum temperature at which the condensate solution exists.  It corresponds to the point on the above plots where
\beq
\left. \frac{\partial v}{\partial T} \right|_{T=T_{c2}} = \infty \, .
\eeq
We see from Figure 1 that lowering the value of $n^{1/3}/\alpha_{\rm em} m$ appears to increase $T_{c2}$ relative to $T_{c1}$.  Lower densities also appear to correspond to larger discontinuities in $v$.

For the charged condensate, as $n^{1/3}$ increases relative to $\alpha_{\rm em} m$ the function $v(T)$ appears to asymptotically approach a single curve that is similar to the curve for the neutral condensate.  The phase transition appears to be nearly very second order with $v$ increasing more-or-less continuously as $T$ decreases.  Even though the phase transition may technically be first order the jump in $v$ becomes negligibly small.

The above contours were drawn using the approximation $\mu = m$ in the condensed phase.  The actual condensate solutions are slightly to the left of these contours, when $\mu = m+O(\alpha_{\rm em} T)$.

The existence of a first order phase transition at low densities can be understood in the following way.  In the low density regime the mass of the bosonic excitation $gv$ becomes comparable to or greater than the transition temperature.   Thus the thermal contributions of this mode to the number density become exponentially suppressed by a factor of $\exp(-gv/T)$ in the condensed phase.  In order to maintain the fixed charge density, $v$ must jump discontinuously so that charge that had been stored in thermal excitations is stored instead in the condensate.  As the density is lowered, $T_c$ continues to decrease relative to $gv$ and the effect grows more pronounced.

\vspace{0.3cm}

\begin{figure}[t]
\begin{center}
\epsfxsize=3in
\epsffile{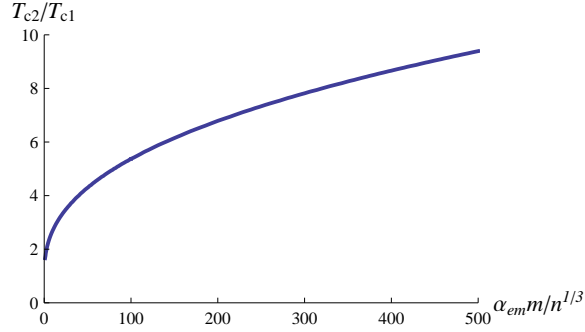}
\end{center}
\caption{Critical temperatures $T_{c2}/T_{c1}$ as a function of $\alpha_{\rm em} m /n^{1/3}$ when $g=2e$.}
\label{fig2}
\end{figure}

Let us now find $T_{c2}$.  In the nonrelativistic limit we can approximate the number density as
\beq
n
 \simeq  mv^2+ \int{\frac{d^3p}{(2 \pi)^3}} \frac{p^2}{2m \omega_-} \frac{1}{\esp^{\beta \omega_-}-1} \, ,
\eeq 
where $\omega_- \simeq \sqrt{p^4/4m^2+g^2 v^2}$.  Here we have taken $\mu \simeq m$ in the condensed phase.  Changing to dimensionless variables
\beq
x \equiv \frac{p^2}{2mT} \, , ~~~~{\rm and}~~~~ y \equiv \frac{gv}{T} \, ,
\eeq
the number density becomes
\beq
n
 \simeq mv^2 +\left(\frac{mT}{2 \pi}\right)^{3/2} \frac{2}{\sqrt{\pi}}  \int_0^\infty{\frac{dx\, x^{3/2}}{\sqrt{x^2+y^2}}}\frac{1}{\esp^{\sqrt{x^2+y^2}}-1} \, .
\eeq 

Let us assume that at the critical temperature $T_{c2}$ the expectation value of the condensate as given by the above expression is roughly the same order of magnitude as it is at zero temperature: $v \sim \sqrt{n/m}$.   Let us also assume that $T_{c2}$ is roughly the same order of magnitude as $T_{c1}$ as given by (\ref{Tc1}).  Then when $n^{1/3} \ll \alpha_{\rm em} \, m$ we have $T_{c2} \ll gv$.  In terms of our dimensionless variables this is $y \gg 1$.  In this regime we can do a low temperature expansion of the above integral \cite{HW2}.  The first two terms of the low temperature expansion give:
\beq
\label{nmum}
n \simeq m v^2+\frac{2^{5/4} \, \Gamma[5/4]}{\sqrt{\pi}}\left( \frac{m T}{2 \pi} \right)^{3/2} \left[ \left(\frac{g v}{T}\right)^{1/4} {\rm Li}_{5/4}\left(\esp^{-gv/T}\right)+\frac{5}{32} \left(\frac{g v}{T}\right)^{-3/4} {\rm Li}_{9/4}\left(\esp^{-gv/T}\right)\right] \, , \nonumber \\
\eeq
where Li is the polylogarithm.

$T_{c2}$ corresponds to the maximum temperature at which this expression can be satisfied for a fixed $n$.  Equivalently, it corresponds to the critical point of the function $v(T)$.  So to find $T_{c2}$ we differentiate both sides of the above expression with respect to $v$ and we set $\partial T/\partial v =0$.  We find that at the critical temperature, $T_{c2}$ and $v$ are related by
\beq
v_{c2} \simeq \sqrt{\frac{n}{m}+\frac{1}{2}\left(\frac{T_{c2}}{g}\right)^2}-\frac{T_{c2}}{g} \, .
\eeq
Substituting this expression for $v$ into (\ref{nmum}) gives an implicit expression for $T_{c2}$ in terms of $n$ and $m$.  

In figure 2 we plot the ratio of $T_{c2}$ to $T_{c1}$ as a function of $\alpha_{\rm em} m/n^{1/3}$, from $\alpha_{\rm em} m/n^{1/3}=1$ to $500$.   Again we set $g=2e$.  As $n$ decreases the ratio $T_{c2}/T_{c1}$ grows.  Since $T_{c1}$ coincides with the critical temperature for the neutral condensate, this implies that for the charged condensate at low densities, condensation can occur at temperatures several times higher than for the neutral condensate.  In high density regimes where $\alpha_{\rm em} m/n^{1/3}  < 1$,  the above approximations break down.  We expect that in this limit the Coulomb energy of the scalars should be small compared to their kinetic energy and thus $T_{c2}/T_{c1}$ should approach $1$.

\section{Conclusions and Discussion}

We have calculated the one-loop grand canonical partition function for scalar electrodynamics at finite temperature and chemical potential.  We considered an electrically neutral system in which the charged scalars reside in a background density of the opposite charge.  Using the background field method we found a result that was gauge condition dependent when the background fields did not satisfy the tree-level equations of motion.  To address this, we required consistency with a Ward identity in order to determine the off-shell thermodynamic potential.

Using this potential we could determine, to lowest order in perturbation theory, the thermal expectation value of the charged scalars $v$ as a function of temperature.  Plotting $v(T)$ for various values of the scalar mass $m$ and number density $n$, we found evidence of a first order phase transition in the low density regime, when $n^{1/3} \lesssim \alpha_{\rm em} m$.  In between two critical temperatures $T_{c1}$ and $T_{c2}$, $v$ jumps discontinuously from zero.  The lower temperature $T_{c1}$ coincides with the critical temperature for a neutral condensate, while $T_{c2}$ can be several times higher in the low density regime.  Thus the phase transition for the charged condensate can occur at higher temperatures than for the neutral condensate.  The strength of the first order phase transition appears to increase with decreasing density.

The critical temperatures $T_{c1}$ and $T_{c2}$ give only lower and upper bounds on the transition temperature.  Finding the exact transition temperature as well as other physically interesting quantities, such as the latent heat involved in the phase transition or the timescale for the transition to occur, requires a more thorough understanding of the energetics of these solutions and their stability or metastability.  Investigation of these questions would entail using the constraint equation $n=-\partial \Omega/\partial \mu$ to ``integrate out" $\mu$ so that the free energy density $F(\mu, v,T) = \Omega(\mu, v,T) +\mu n\rightarrow F(n,v,T)$.   This is analogous to our procedure for the zero-temperature potential in equation (\ref{V}).

\vspace{0.3cm}

\begin{figure}[t]
\begin{center}
\epsfxsize=3in
\epsffile{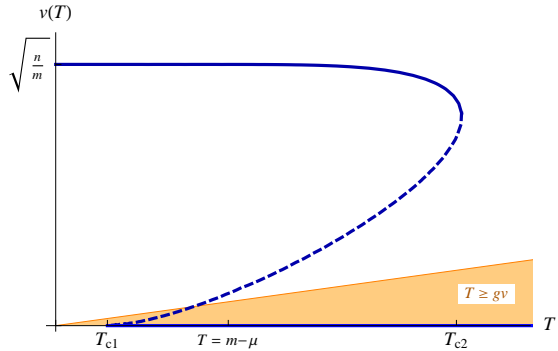}
\end{center}
\caption{Region of validity of the perturbative expansion}
\label{fig3}
\end{figure}

The above conclusions were based on a one-loop calculation.  It is reasonable to consider the effects of higher order corrections in perturbation theory.  In thermal field theory at high temperatures, the perturbative expansion can break down, even for weakly coupled theories.  This can occur when the temperature of the system is greater than the masses of the particles running in the loops.  For theories with massless particles, this breakdown is often signaled by the appearance of infrared divergences at higher loops.

For the system described here, because the gauge symmetry is spontaneously broken in the condensed phase, the gauge bosons are massive.  Thus one can compute to any order in perturbation theory without encountering infrared divergences.  Moreover, in the condensate, as long as the temperature is sufficiently below the gauge boson mass $gv$ and the mass of the remaining scalar degree of freedom $2m$, the loop expansion parameter should remain small.  In the uncondensed phase, the massless photons interact with each other only through the massive scalars, the lightest of these scalars having mass $m-\mu$.  Thus as long as $m-\mu \gg T$ our considerations in the uncondensed phase should be valid.  

For a second order phase transition or for a weakly first order phase transition these criteria will be violated near the phase transition point, where $gv \rightarrow 0$ and $\mu \rightarrow m$.  Thus perturbation theory can not be considered reliable in the vicinity the phase transition.  However, for a strongly first order phase transition, which is precisely the interest of this work, this region is avoided.  In the low density regime the condensate solution doesn't pass near $gv =0$; the temperature of the system is always small compared to the masses of the excitations in the condensate.  Thus higher order corrections to the results found above should be small.

To illustrate this we plot $v(T)$ in figure 3 for the values of $m$ and $n$ that are relevant for helium-core white dwarf stars ($m \simeq 4 \times 10^4 \, n^{1/3}$).  The orange region corresponds to where $T \geq gv$ and thus perturbation theory can be unreliable.  The condensate solution which is the upper solid blue curve is well outside this region.  We also mark the point where, when $v=0$, $T = m-\mu$.  To the left of this point, i.e., when $T > m-\mu$, calculations of, say, the free energy density in the {\it uncondensed} phase using the expressions obtained above should not to be trusted.  This may place some limitations on calculating the exact transition temperature, if the phase transition occurs in this region.  However, it is possible that methods such as resummation can be applied in this region to obtain more reliable results.

\vspace{0.3cm}

Let us end with some comments on physical applications.  In helium-core white dwarf stars the number density of the helium nuclei is such that $n^{1/3} < \alpha_{\rm em} m$.  Thus based on the arguments given above, we might expect the condensation of these nuclei to be a first order phase transition.  We may also expect the condensation temperature for the helium nuclei to be somewhat greater than that for neutral bosons.  The latent heat associated with the phase transition could potentially delay the cooling of the star.  For a carbon-core white dwarf star of mass $\sim M_\odot$ which crystallizes, the latent heat associated with crystallization can increase the classical cooling time by a factor of $\sim 1.6$ \cite{LVH}.  However, one of the main features of a condensed-core white dwarf is that it cools much more rapidly than white dwarfs with uncondensed or crystallized cores \cite{GGDP}.  It is possible that any contribution to the cooling history coming from the latent heat will be overwhelmed by the rapid cooling in the condensed phase.

In this work we treated the background charge density as static and classical.  For the system of nuclei and electrons in the cores of white dwarfs, this is not a good approximation.  The electrons, i.e. the background charge density, form a degenerate fermi gas with gapless excitations.  Due to these excitations, the electrons could play a significant role in the phase transition.

\vspace{0.3cm}

\begin{center}
{\bf   Acknowledgments}
\end{center}

I am thankful to Marcus Berg, T. H. Hansson and Fawad Hassan for stimulating conversations and especially to Gregory Gabadadze for conversations and comments on the draft.  This work was supported by the Swedish Research Council (VR) through the Oskar Klein Centre.

\end{document}